\documentstyle[psfig]{article}
\begin{document}
\title{Numerical Study of Cosmological Singularities}
\author{Beverly K. Berger, David Garfinkle \\
Physics Department, Oakland University, Rochester, MI 48309 USA \\
\\
Vincent Moncrief \\
Physics Department, Yale University, New Haven, CT 06520 USA}

\maketitle
\begin{abstract}
The spatially homogeneous, isotropic Standard Cosmological Model appears to 
describe our Universe reasonably well. However, Einstein's equations allow a 
much larger class of cosmological solutions. Theorems originally due to Penrose 
and Hawking predict that all such models (assuming reasonable matter properties)
will have an initial singularity. The nature of this singularity in generic 
cosmologies remains a major open question in general relativity. Spatially 
homogeneous but possibly anisotropic cosmologies have two types of 
singularities: (1) velocity dominated---(reversing the time direction) the 
universe evolves to the singularity with fixed anisotropic collapse rates ; 
(2) Mixmaster---the anisotropic collapse rates change in a deterministically
chaotic way. Much less is known about spatially inhomogeneous universes.
Belinskii, Khalatnikov, and Lifshitz (BKL) claimed long ago that a generic
universe would evolve toward the singularity as a different Mixmaster universe
at each spatial point. We shall report on the results of a program to test the
BKL conjecture numerically. Results include a new algorithm to evolve
homogeneous Mixmaster models, demonstration of velocity dominance and
understanding of evolution toward velocity dominance in the plane symmetric
Gowdy universes (spatial dependence in one direction), demonstration of
velocity dominance in polarized U(1) symmetric cosmologies (spatial dependence
in two directions), and exploration of departures from velocity dominance in
generic U(1) universes. 
\end{abstract}
\section{Introduction}
We shall describe a series of numerical studies of the nature of singularities
in cosmological models. Since the interiors of black holes can be described
locally as cosmological models, it is possible that our methods and results may
be useful to the participants in this conference.

The generic singularity in spatially homogeneous cosmologies is reasonably well
understood. The approach to it asymptotically falls into two classes. The
first, called asymptotically velocity term dominated (AVTD)
\cite{ber-els,ber-IM}, refers to a cosmology that approaches the
Kasner (vacuum, Bianchi I) solution \cite{ber-kasner} as $\tau \to
\infty$. (Spatially homogeneous universes can be described as a sequence of
homogeneous spaces labeled by
$\tau$. Here we shall choose $\tau$ so that $\tau = \infty$ coincides with the
singularity.) An example of such a solution is the vacuum Bianchi II model
\cite{ber-taub} which begins with a fixed set of Kasner-like anisotropic
expansion rates, and, possibly, makes one change of the rates in a prescribed
way (Mixmaster-like bounce) and then continues to $\tau = \infty$ as a fixed
Kasner solution. In contrast are the homogeneous cosmologies which display
Mixmaster dynamics such as vacuum Bianchi VIII and IX
\cite{ber-BKL,ber-misner,ber-halpern} and Bianchi VI$_0$ and Bianchi I with a
magnetic field \cite{ber-LKW,ber-bkb1,ber-leblanc}. Jantzen
\cite{ber-jantzen} has discussed other examples. Mixmaster dynamics describes an
approach to the singularity which is a sequence of Kasner epochs with a
prescription, originally due to Belinskii, Khalatnikov, and Lifshitz (BKL)
\cite{ber-BKL}, for relating one Kasner epoch to the next. Some of the Mixmaster
bounces (era changes) display sensitivity to initial conditions one usually
associates with chaos and in fact Mixmaster dynamics is chaotic
\cite{ber-cornish}. The vacuum Bianchi I (Kasner) solution is distinguished
from the other Bianchi types in that the spatial scalar curvature $^3\!R$,
(proporional to) the minisuperspace (MSS) potential \cite{ber-misner,ber-ryan},
vanishes identically. But
$^3\!R$ arises in other Bianchi types due to spatial dependence of the metric
in a coordinate basis. Thus an AVTD singularity is also characterized as a
regime in which terms containing or arising from spatial derivatives no longer
influence the dynamics. This means that the Mixmaster models do not have an
AVTD singularity since the influence of the spatial derivatives (through the
MSS potential) never disappears---there is no last bounce. 

In the late 1960's, BKL claimed to show that singularities in generic solutions
to Einstein's equations are locally of the Mixmaster type \cite{ber-BKL}. This
means that each point of a spatially inhomogeneous universe could collapse to
the singularity as the Mixmaster sequence of Kasner models. (It has been argued
that this could generate a fractal spatial structure
\cite{ber-montani,ber-belinskii,ber-KK}.) In contrast, each point of a cosmology
with an AVTD singularity evolves asymptotically as a fixed Kasner model.
Although the BKL result is controversial \cite{ber-bt}, it provides a hypothesis
for testing. Our ultimate objective is to test the BKL conjecture numerically.

\section{Numerical Methods}
The work reported here was performed by using symplectic ODE and PDE solvers
\cite{ber-fleck,ber-vm83}. While other numerical methods may be used to solve
Einstein's equations for the models discussed here, symplectic methods have
proved extremely advantageous for Mixmaster models and have also worked quite
well in the Gowdy plane symmetric and polarized $U(1)$ symmetric cosmologies
\cite{ber-bkb96,ber-bkbvm,ber-bggm,ber-bkbvm2}. Consider a system with one
degree of freedom described by $q(t)$ and its canonically conjugate momentum
$p(t)$ with a Hamiltonian
\begin{equation}
\label{ber-1dofH}
H = {{p^2} \over {2m}} + V(q) = H_K + H_V.
\end{equation}
Note that the subhamiltonians $H_K$ and $H_V$ separately yield equations of
motion which are exactly solvable no matter the form of $V$. Variation of $H_K$
yields $\dot q = p/m$, $\dot p = 0$ with solution
\begin{equation}
\label{ber-HKsoln}
p(t + \Delta t) = p(t) \quad, \quad q(t + \Delta t) = q(t) + {{p(t)} \over m}
\Delta t.
\end{equation}
Variation of $H_V$ yields $\dot q = 0$, $\dot p = -dV/dq$ with solution
\begin{equation}
\label{ber-HVsoln}
q(t + \Delta t) = q(t) \quad, \quad p(t + \Delta t) = p(t) - \left. {{{dV}
\over {dq}}} \right|_t \, \Delta t.
\end{equation}
Note that the absence of momenta in $H_V$ makes (\ref{ber-HVsoln}) exact for any
$V(q)$. One can then demonstrate that to evolve from $t$ to $t + \Delta t$ an
evolution operator ${\cal U}_{(2)}(\Delta t)$ can be constructed from the
evolution sub-operators ${\cal U}_K(\Delta t)$ and ${\cal U}_V(\Delta t)$
obtained from (\ref{ber-HKsoln}) and (\ref{ber-HVsoln}). One can show that
\cite{ber-fleck}
\begin{equation}
\label{ber-Uprescription}
{\cal U}_{(2)}(\Delta t) = {\cal U}_K(\Delta t/2)\,{\cal U}_V(\Delta
t)\,{\cal U}_K(\Delta t/2)
\end{equation}
reproduces the true evolution operator through order $(\Delta t)^2$. Suzuki has
developed a prescription to represent the full evolution operator to arbitrary
order \cite{ber-suzuki}. For example
\begin{equation}
\label{ber-4thorderU}
{\cal U}_{(4)}(\Delta t) ={\cal U}_{(2)}(s\Delta t)\,{\cal U}_{(2)}[(1-2s)\Delta
t]\,{\cal U}_{(2)}(s\Delta t)
\end{equation}
where ${s} = 1/(2 - {2^{1/3}})$. The advantage of Suzuki's approach is
that one only needs to construct ${\cal U}_{(2)}$ explicitly. ${\cal U}_{(2n)}$
is then constructed from appropriate combinations of ${\cal U}_{(2n-2)}$.

The generalization of this method to $N$ degrees of freedom and to fields is
straightforward. In the latter case, $V[\vec q(t)] \to V[\vec
q(\vec x,t)]$ so that $dV/dq$ becomes the functional derivative $\delta V /
\delta q$. On the computational spatial lattice, the derivatives that are
obtained in the expression for the functional derivative must be represented in
differenced form. We note that, to preserve $n$th order accuracy in time,
$n$th order accurate spatial differencing is required. Some discussion of this
has been given elsewhere
\cite{ber-bkbvm}.

\section{Application to Mixmaster Dynamics}

(Diagonal) Bianchi Class A cosmologies \cite{ber-ryan} are described by the
metric
\begin{equation}
\label{bianchimetric}
ds^2 = -\, e^{3 \Omega} \,dt^2 \,+\, \left( e^{2\beta} \right)_{ij} \,
\sigma^i\,\sigma^j
\end{equation}
where $\beta_{ij} = {\rm diag}
(-\,2\beta_+,\,\beta_+\,+\,\sqrt{3}\,\beta_-,\,\beta_+ \,-\,\sqrt{3}\,
\beta_-)$,
$d\sigma^i = C^i_{jk} \sigma^j \wedge \sigma^k$ defines the Bianchi type and
$t$ is the BKL time coordinate.  Einstein's equations can be obtained by
variation of the Hamiltonian
\begin{equation}
\label{ber-h0homog}
2H = -\,p_\Omega^2 + p_+^2 +p_-^2 + V(\beta_+, \beta_-, \Omega).
\end{equation}
The logarithmic anisotropic scale factors $\alpha$, $\zeta$, and $\gamma$ are
given in terms of the logarithmic volume $\Omega$ and anisotropic shears
$\beta_\pm$ as
\begin{eqnarray}
\label{ber-variables}
\alpha &=& \Omega - 2 \beta_+ \nonumber , \\
\zeta  &=& \Omega + \beta_+ + \sqrt{3}\, \beta_- \nonumber, \\
\gamma &=&  \Omega + \beta_+ - \sqrt{3}\, \beta_- .
\end{eqnarray}
The momenta $p_\Omega$, $p_\pm$ are canonically conjugate to $\Omega$,
$\beta_\pm$ respectively. For vacuum Bianchi IX or magnetic Bianchi VI$_0$,
the MSS potential $V$ has the form \cite{ber-bkb96}
\begin{equation}
\label{ber-MSSV}
V = {c^2 \, e^{2 b \alpha}} \, + \,  {e^{4 \zeta}} \, + \,  {e^{4
\gamma}}
\; -
\; 2 \, \left ( a \, {e^{2 (\alpha + \zeta )}} \, + \, a \, {e^{2 (\alpha +
\gamma )}}
\, + \, d \, {e^{2 (\zeta + \gamma )}} \right )   \; \; \; .
\end{equation}
Here $a = 1$, $b=2$, $c=1$, and
$d = 1$ for vacuum Bianchi Type IX, while  $a = 0$, $b=1$, $c = {\sqrt \xi} $,
and $ d = -1$ for magnetic Bianchi Type VI$_0$,   
and $\xi $ is a constant that depends on the strength of the magnetic field. In
these models, the singularity occurs at $\tau = - \Omega = \infty$. From
(\ref{ber-variables}) and (\ref{ber-MSSV}), we see that $V \to 0$ as $\tau \to
\infty$ unless one of
$\alpha$,
$\zeta$, or $\gamma$ is $\approx 0$. When the potential itself is small ($V
\approx 0$), (\ref{ber-h0homog}) describes a ``free particle'' in MSS and is in
fact approximately the Kasner solution.
\begin{figure}[bth]
\begin{center}
\setlength{\unitlength}{1cm}
\makebox[11.7cm]{\psfig{file=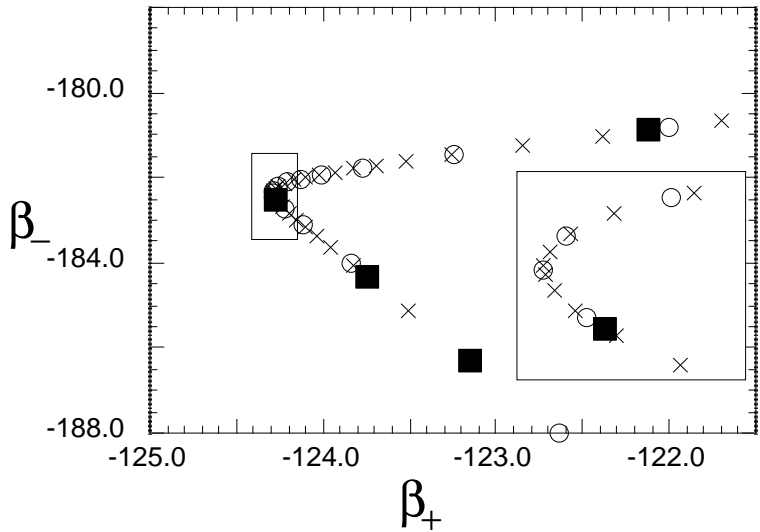,width=10cm}}
\caption[Compare Algorithms]
{\protect \label{ber-fig1}
Comparison of algorithms for Mixmaster dynamics. A typical Mixmaster
bounce is shown in the anisotropy plane. Crosses indicates every 10th point on a
4th order Runge-Kutta evaluation of the trajectory, while circles indicate every
10th point for a 6th order standard symplectic evaluation. The filled squares
indicate {\it every} point using the new algorithm. The inset shows the details
closest to the bounce. Note that the new algorithm does not require a point at
the apex of the bounce.}
\end{center}
\end{figure}

\begin{figure}[bth]
\begin{center}
\setlength{\unitlength}{1cm}
\makebox[11.7cm]{\psfig{file=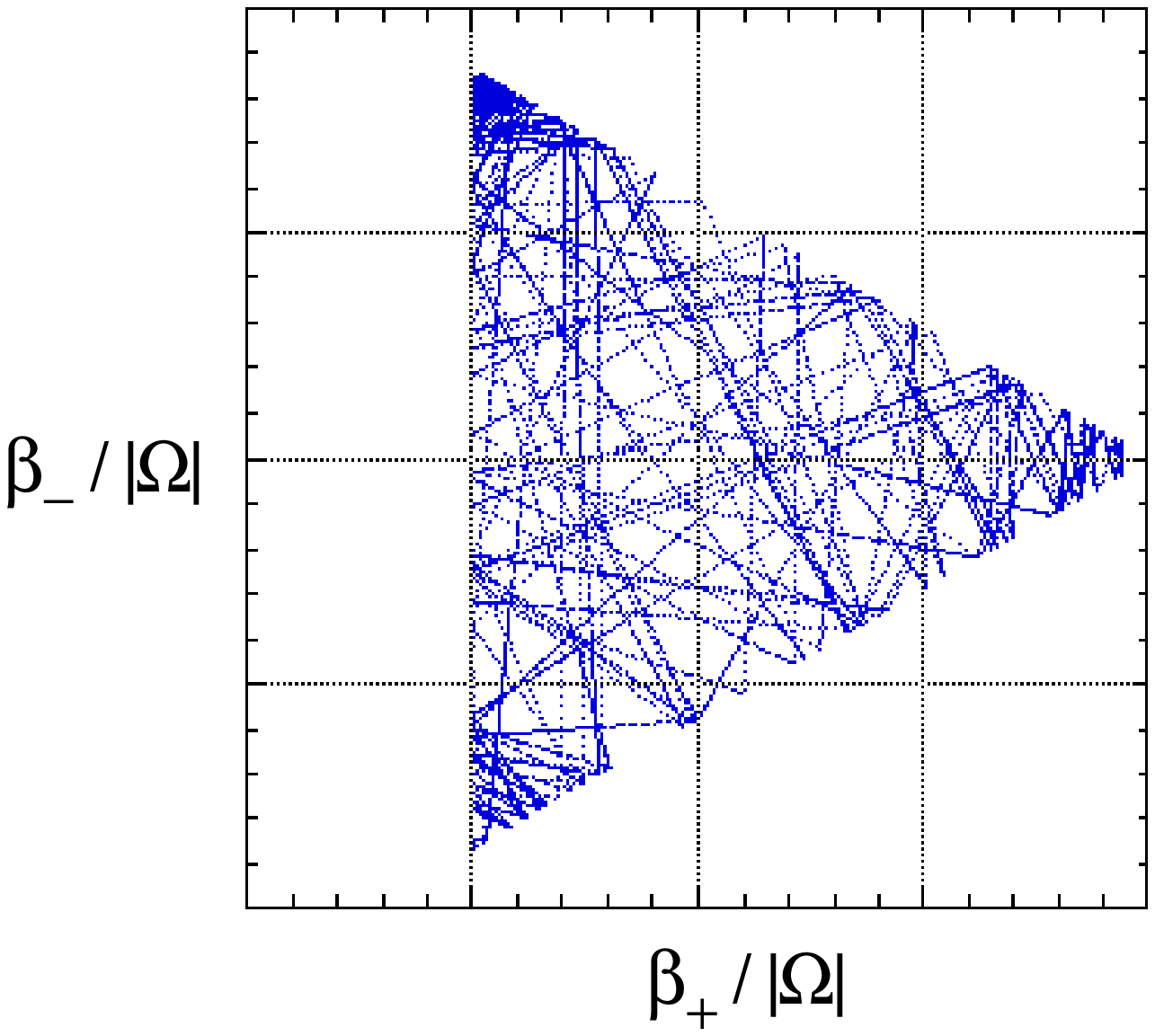,width=10cm}}
\caption[Mixmaster trajectory]
{\protect \label{ber-fig2}
A typical trajectory consisting of 268 Mixmaster bounces is shown projected onto
the anisotropy plane. In terms of the rescaled variables $\beta_+ / |\Omega|$
and $\beta_- / |\Omega|$, the walls of the MSS potential and thus the bounce
sites are at fixed locations.}
\end{center}
\end{figure}

The standard algorithms for solving ODE's \cite{ber-nr} often employ an adaptive
time step. The idea is to take large time steps where nothing much happens (e.g.
\/in the Kasner regime) while taking shorter time steps when the forces are
large (e.g. \/at a bounce). Unfortunately, in Mixmaster dynamics, the duration
of the Kasner epochs increases exponentially in $\tau$ as $\tau \to \infty$
\cite{ber-sinai} while the duration of the bounce itself is in some sense fixed
\cite{ber-bkb91}. This means that, although huge time steps may be taken in the
Kasner segments, the time step must become very small at the bounces. Ideally,
one would like the time step also to grow exponentially but this cannot be done
with standard approaches. Thus, with a great deal of effort, standard methods
can yield about 30 bounces for $0 < \tau < 10^8$ in about an hour on a
supercomputer. The application of the symplectic method to Mixmaster dynamics
is straightforward. From (\ref{ber-h0homog}), we have 
\begin{equation}
\label{ber-kasnersplit}
H_K = - \,p_\Omega^2 + p_+^2 + p_-^2 \quad ; \quad H_V = V(\Omega, \beta_+,
\beta_-).
\end{equation}
With an adaptive step size and a 6th order version of
(\ref{ber-4thorderU}), one can do slightly better (approximately a factor of
three fewer steps) than 4th order Runge-Kutta. However, it is well known that a
bounce off an exponential wall---the Bianchi II (or Taub \cite{ber-taub})
cosmology---is exactly solvable. If we first identify the dominant exponential
wall (say $e^{4 \alpha}$), then we find that the symplectic algorithm works for
a different split of the Hamiltonian into two subhamiltonians
\cite{ber-bkb96}. Let
\begin{equation}
\label{ber-h0taub}
H = H_1 + H_2
\end{equation}
where (e.g. \/for Bianchi IX)
\begin{equation}
\label{ber-h1h2taub}
H_1 = - \,p_\Omega^2 + p_+^2 + p_-^2 + e^{4 \Omega - 8 \beta_+} \quad ; \quad
H_2 = H_V - e^{4 \Omega - 8 \beta_+}.
\end{equation}
Variation of $H_2$ is exactly solvable as before, while variation of $H_1$
yields equations with solution
\begin{eqnarray}
\label{ber-h1soln}
p_y(t + \Delta t) &=& p_y(t) \quad; \quad p_-(t + \Delta t) = p_-(t)
\quad ; \nonumber \\
\nonumber \\
y(t + \Delta t) &=& y(t) - 6 p_y(t) \Delta t \quad ; \quad \beta_-(t + \Delta t)
= \beta_-(t) + 2 p_-(t) \Delta t  \quad ; \nonumber \\
\nonumber \\
\alpha(t + \Delta t) &=& - {1 \over 2} \ln \left[ {1 \over E} \cosh 4 \sqrt{3}\,
(t + \Delta t - t_0) \right] \quad ; \quad 6 p_\alpha = {{d \alpha} \over {dt}}
\end{eqnarray}
where $y = - 2 \Omega + \beta_+$ and $E^2 = 3 p_y^2 - p_-^2$.
This new symplectic algorithm provides an enormous advantage because the bounce
is built in. The time step can grow exponentially. In a few minutes on a
desktop computer, one can obtain, e.g., 268 bounces for $8 < \tau <
10^{61.5}$.   Fig.~\ref{ber-fig1} shows a comparison of the new and standard
algorithms while Fig.~\ref{ber-fig2} shows a typical trajectory. Since $H_1$ is
actually the exact Hamiltonian almost all the time, the accuracy of the method
is much higher than the formal 6th order---we achieve machine precision. 

The BKL parameter $u$ characterizes the angle in the anisotropy plane of the
Kasner epoch trajectory. Between the $n$th and $n+1$st Kasner epochs, we have
\begin{equation}
\label{ber-umap}
u_{n+1}=\left\{ \matrix{u_n-1\quad \quad \quad ;\quad \quad 2\le u_n<\infty
\hfill\cr
\cr
(u_n-1)^{-1}\quad ;\quad \quad 1\le u_n\le 2\hfill\cr} \right. .
\end{equation}
All numerical studies \cite{ber-BKL,ber-moser,ber-rugh,ber-bkb91} and a variety
of analytic arguments \cite{ber-rendall,ber-bkb96} have shown that one expects
(\ref{ber-umap}) to become ever more valid as $\tau \to
\infty$. With our algorithm  in a double or quadruple precision code, one can
evolve until deviations from (\ref{ber-umap}) disappear and then use the
predicted values of $u$ as a test of the accuracy of the code. For example
\cite{ber-bkb96}, one discovers that the Hamiltonian constraint 
((\ref{ber-h0homog}) with $H = 0$) must be enforced although not necessarily at
every time step. 

For our studies of spatially inhomogeneous cosmologies, it is important to keep
in mind that

(1) between bounces Mixmaster dynamics looks like the AVTD Kasner solution;

(2) in a fixed time variable such as $\tau$, the time between bounces increases
exponentially as $\tau \to \infty$;

(3) extraordinary accuracy can be achieved by symplectic methods when one of
the subhamiltonians is dominant;

(4) enormous gains in accuracy and computational speed can be made by using a
``custom-designed'' treatment of the bounce.

\section{The Gowdy Test Case}
As the simplest example of a spatially inhomogeneous cosmology, we consider the
plane symmetric vacuum Gowdy universe on $T^3 \times R$
\cite{ber-gowdy,ber-bkb74} described by the metric
\begin{eqnarray}
\label{ber-gowdymetric}
ds^2&=&e^{{{-\lambda } \mathord{\left/ {\vphantom {{-\lambda 
} 2}} \right. \kern-\nulldelimiterspace} 2}}\,e^{{\tau  \mathord{\left/ 
{\vphantom {\tau  2}} \right. \kern-\nulldelimiterspace} 2}}\,(-\,e^{-2\tau 
}\,d\tau ^2\,+\,d\theta ^2)\nonumber \\
\nonumber \\
 &  &+\,e^{-\tau }\,[e^P\,d\sigma ^2\,+\,2\,e^P\,Q\,d\sigma \,d\delta
\,+\,(e^P\,Q^2\,+\,e^{- P})\,d\delta ^2]
\end{eqnarray}
where the background $\lambda$ and amplitudes $P$ and $Q$ of the $+$ and
$\times$ polarizations of gravitational waves are functions of $\tau$
and $0 \le \theta \le 2 \pi$ and periodic in $\theta$. There is a curvature
singularity at $\tau =
\infty$ \cite{ber-bkb74,ber-IM,ber-vm81}. The polarized case ($Q = 0$) has been
shown to have an AVTD singularity \cite{ber-IM}. The generic case has been
conjectured to be AVTD (except perhaps at a set of measure zero) \cite{ber-bm1},
which we have verified to the extent possible numerically
\cite{ber-bkbvm,ber-bkb94a,ber-bkb94b,ber-bkb97a,ber-bkb97b,ber-bkb97c,ber-bggm}.
A claim has been made that this model does not have an AVTD singularity
\cite{ber-stewart} which we believe to be incorrect \cite{ber-bgm}.

This model is especially attractive as a test case because Einstein's equations
in our variables split into dynamical equations for the wave amplitudes
(where $,_a = \partial / {\partial a}$):
\begin{equation}
\label{ber-gowdywaveP}
P,_{\tau \tau }-\;e^{-\kern 1pt2 \tau }P,_{\theta \theta }\,-\,e^{2P}\left( 
{Q,_\tau ^2-\;e^{-\kern 1pt2\tau }Q,_\theta ^2} \right) = 0,
\end{equation}

\begin{equation}
\label{ber-gowdywaveQ}
  Q,_{\tau \tau }-\;e^{-\kern 1pt2\tau }Q,_{\theta \theta }+\,2\,\left( 
{P,_\tau Q,_\tau ^{}-\;e^{-\kern 1pt2\tau }P,_\theta Q,_\theta ^{}} 
\right) = 0
\end{equation}
while the Hamiltonian and momentum constraints become respectively first order
equations for $\lambda$:
\begin{equation}
\label{ber-gowdyh0}
\lambda ,_\tau -\;[P,_\tau ^2+\;e^{-2\tau }P,_\theta ^2+\;e^{2P}(Q,_\tau 
^2+\;e^{-2\tau }\,Q,_\theta ^2)]=0,
\end{equation}

\begin{equation}
\label{ber-gowdyhq}
\lambda ,_\theta -\;2(P,_\theta P,_\tau +\;e^{2P}Q,_\theta Q,_\tau )=0.
\end{equation}
Thus two problematical aspects of numerical relativity---preservation of the
constraints and solution of the initial value problem---become trivial
\cite{ber-gowdy}. The wave equations (\ref{ber-gowdywaveP}) and
(\ref{ber-gowdywaveQ}) may be obtained by variation of the Hamiltonian
\begin{eqnarray}
\label{ber-gowdywaveh}
H&=&{\textstyle{1 \over 2}}\int\limits_0^{2\pi } {d\theta 
\,\left[ {\pi _P^2+\kern 1pt\,e^{-2P}\pi _Q^2} \right]}\nonumber \\
  &+&{\textstyle{1 \over 2}}\int\limits_0^{2\pi } {d\theta \,\left[ {e^{-
2\tau }\left( {P,_\theta ^2+\;e^{2P}Q,_\theta ^2} \right)} 
\right]}=H_K+H_V 
\end{eqnarray}
where $\pi_P$ and $\pi_Q$ are canonically conjugate to $P$ and $Q$ respectively.
Variation of $H_K$ yields the AVTD solution
\begin{eqnarray}
\label{ber-avtdeq}
P&=&\ln |\mu|\,+\,v(\tau -\tau _0)\,+\,\ln [1+e^{-2v(\tau -\tau _0)}] \to v\tau
\quad {\rm as}\;\tau \to \infty , \nonumber \\
\nonumber \\
Q&=&Q_0 \,+\,{1 \over {\mu}}{{e^{-2v(\tau -\tau _0)}}\over{(1+e^{-2v(\tau -\tau
_0)})}}
\quad \quad \quad \quad \quad \ \    \;\to Q_0\;\quad {\rm as}\;\tau \to
\infty , \nonumber \\
\nonumber \\
\pi _P&=&v{{\,(1-e^{-2v(\tau -\tau _0)})} \over {\,(1+e^{-2v(\tau -\tau
_0)})}}\quad \quad \quad \quad \quad  \quad \quad \quad \quad\to v\quad
\;\;{\rm as}\;\tau \to \infty, \nonumber \\
\nonumber \\
\pi _Q&=&-2\mu \,v 
\end{eqnarray}
and is thus exactly solvable in terms of four functions of $\theta$: $\mu$, $v
> 0$, $Q_0$, and $\tau_0$.
$H_V$ is also (trivially) exactly solvable so that symplectic methods can be
used
\cite{ber-bkbvm}.
\begin{figure}[bth]
\begin{center}
\setlength{\unitlength}{1cm}
\makebox[11.7cm]{\psfig{file=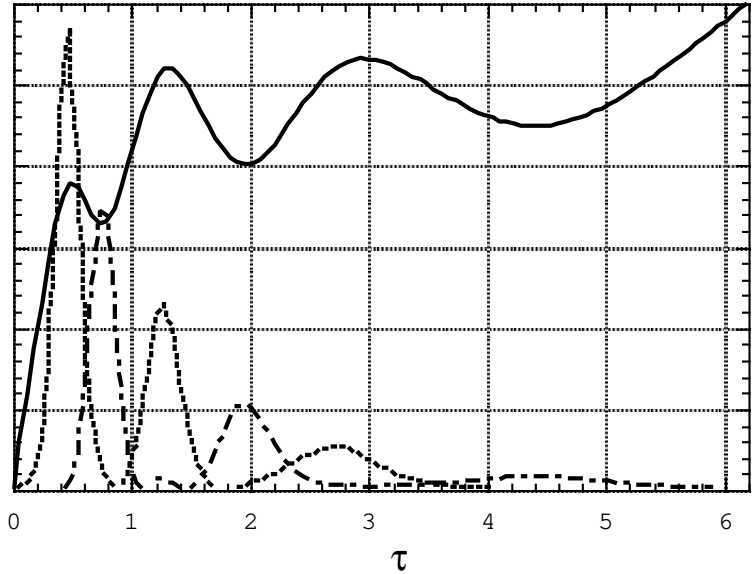,width=10cm}}
\caption[Gowdy potentials]
{\protect \label{ber-fig3}
$P$ (solid curve), $V_1$ (dash-dotted curve), and $V_2$ (dashed curve) vs
$\tau$ at a fixed value of $\theta$. Note that the slope of $P$, $P,_\tau$,
decreases after each interaction with $V_2$ while $P,_\tau$ goes from negative
to positive after each interaction with $V_1$. A continuation of this graph in
$\tau$ would show that
$V_1$ and $V_2$ have permanently died off and that $P$ continues to increase
with fixed positive slope $P,_\tau < 1$.}
\end{center}
\end{figure}

If the singularity is AVTD, one would expect the spatial derivative terms to go
to zero exponentially as $\tau \to \infty$. However, if $P \to v\tau$, the term
\begin{equation}
\label{ber-V2}
V_2 = e^{-2\tau + 2 P} Q,_\theta^2
\end{equation}
in (\ref{ber-gowdywaveP}) would grow rather than decay if $v > 1$. Thus
Grubi\u{s}i\'{c} and Moncrief (GM) \cite{ber-bm1} conjectured that, in a
generic Gowdy model as $\tau
\to \infty$,
$0 \le v < 1$ except perhaps at isolated spatial points. If we consider generic
initial data---e.g. \/$P = 0$, $\pi_P = v_0 \cos \theta$, $Q = \cos \theta$,
$\pi_Q=0$---then we must ask how a generic Gowdy solution evolves toward the
AVTD solution at each spatial point and how an initial $P,_\tau > 1$ or
$P,_\tau < 0$ is brought into the conjectured range. Typically, either $V_2$ or 
\begin{equation}
\label{ber-V1}
V_1 = \pi_Q^2 \, e^{-2P}
\end{equation}
(where $\pi_Q = e^{2P} \,Q,_\tau$) will dominate (\ref{ber-gowdywaveP}) to
yield either 
\begin{equation}
\label{ber-k1}
P,_\tau^2 \,+\, \pi_Q^2 \, e^{-2P} \approx \kappa_1^2
\end{equation}
or
\begin{equation}
\label{ber-k2}
Z,_\tau^2 \,+\, Q,_\theta^2 \, e^{2 Z} \approx \kappa_2^2
\end{equation}
where $Z = P \,-\,\tau$. If $P,_\tau < 0$, an interaction with $V_1$ will occur
to drive $P,_\tau$ to
$-P,_\tau$ to yield $P,_\tau > 0$. If $P,_\tau > 1$, an interaction with $V_2$
will occur to drive $(P,_\tau -1)$ to $- (P,_\tau -1)$. If this yields $P,_\tau
< 0$, a second interaction with $V_1$ will occur, etc. When $|P,_\tau | < 1$,
$V_2$ disappears so that, after a possible final interaction with $V_1$, $0 \le
P,_\tau < 1$ forever. A typical sequence of bounces is shown in
Fig.~\ref{ber-fig3}. 
\begin{figure}[bth]
\begin{center}
\setlength{\unitlength}{1cm}
\makebox[11.7cm]{\psfig{file=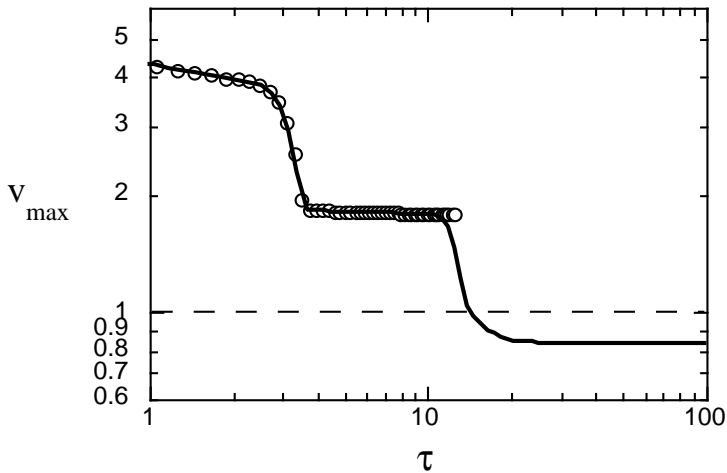,width=10cm}}
\caption[$v_{max}$ vs $\tau$]
{\protect \label{ber-fig4}
Plot of $v_{max}$ vs $\tau$. The maximum value of $v$ is found for two 
simulations with 3200 (solid line) and 20000 spatial grid points (circles)
respectively. The horizontal line indicates $v = 1$. Continuation in $\tau$ of
the finer resolution simulation would show $v_{max} > 1$ for a much longer
$\tau$ with $v_{max} < 1$ eventually.}
\end{center}
\end{figure}
Fig.~\ref{ber-fig4} shows the maximum value of $v$ on the spatial grid vs
$\tau$. First we see that high values of $\tau$ can be reached at which $0 \le
v < 1$ everywhere.
\begin{figure}[bth]
\begin{center}
\setlength{\unitlength}{1cm}
\makebox[11.7cm]{\psfig{file=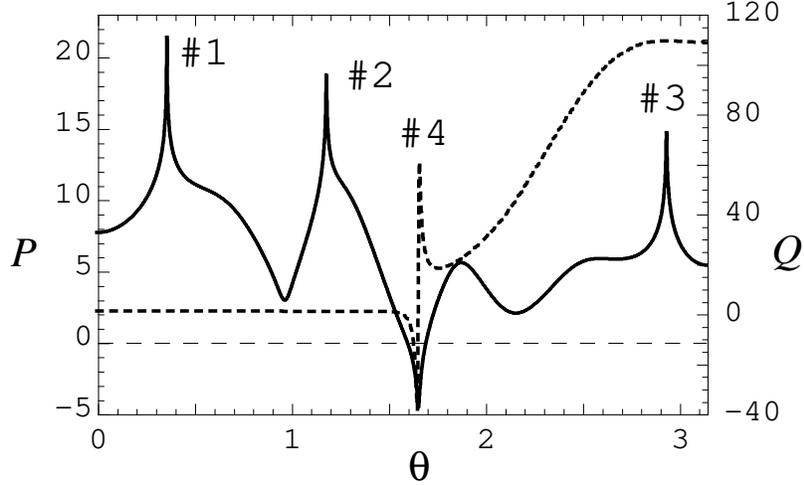,width=11cm}}
\caption[Gowdy $P$ and $Q$ Details]
{\protect \label{ber-fig5}
$P$ (solid line) and $Q$ (dashed line) vs $\theta$ at $\tau = 12.4$ for 
the initial data set given here with $v_0 = 5$ for $0 \le \theta \le \pi$ for
a simulation containing 20000 spatial grid points in the interval $[0,2 \pi]$.
The numbers on the graph refer to the most interesting features. Peaks \#1,
\#2,  and \#3 in $P$ are essentially the same in that they occur
where $Q,_{\theta} \approx 0$.
Peak \#4 shows an apparent discontinuity in $Q$ where 
$\pi_Q \approx 0$.}
\end{center}
\end{figure}

Non-generic behavior can occur at isolated spatial points where either $\pi_Q$
or $Q,_\theta$ vanishes. In the former case, the absence of $V_1$ where $\pi_Q
= 0$ and its flatness where $\pi_Q \approx 0$ allow $P,_\tau$ and thus $P$ to
remain negative for a long time. Since $Q,_\tau = \pi_Q \,e^{-2P}$, $Q$ will
grow exponentially in opposite directions on either side of the points where
$\pi_Q = 0$ producing a characteristic apparent discontinuity. On the other
hand, if $Q,_\theta \approx 0$, $P,_\tau$ can remain large for a long time
causing a spiky feature in $P$. Both types of features sharpen and narrow with
time. The features and their association with non-generic points are shown in
Fig.~\ref{ber-fig5}. The presence of this non-generic behavior at isolated
spatial points leads to a dependence of simulation results on the spatial
resolution. The finer the spatial resolution, the closer will a grid point be
to the non-generic point. Near these non-generic points, the generic process of
approach to $0 \le P,_\tau < 1$ will occur but slowly since either $\pi_Q$ or
$Q,_\theta \approx 0$. The closer one is to a non-generic site, the longer this
process will take. Thus a finer resolution code will have narrower spiky
features at which it takes longer for $P,_\tau$ to move into the range $[0,1)$.
Some evidence for this is seen in Fig.~\ref{ber-fig4} where the finer spatial
resolution simulation diverges from the coarser one and will be considered in
detail elsewhere
\cite{ber-bggm}.

Finally, we note that $V_2$ is analogous to the MSS potential. One may consider
then application of the new Mixmaster algorithm to the Gowdy Hamiltonian
(\ref{ber-gowdywaveh}) with $H = H_1 + H_2$ for 
\begin{equation}
\label{ber-gowdyh1}
H_1 = {\textstyle{1 \over 2}} \oint \, d\theta \left(\pi_P^2 + e^{-2\tau +
2P}Q,_\theta^2
\right)
\end{equation}
and
\begin{equation}
\label{ber-gowdyh2}
H_2 = {\textstyle{1 \over 2}} \oint \, d\theta \left(\pi_Q^2 \,e^{-2P} +
e^{-2\tau}P,_\theta^2
\right).
\end{equation}
These yield the exact solutions
\begin{eqnarray}
\label{ber-newh1soln}
Q(\tau + \Delta \tau) &=& Q(\tau) , \nonumber \\
\nonumber \\
\pi_Q(\tau + \Delta \tau) &=& \xi + \left[ {{\kappa} \over
{Q(\tau),_\theta}} \, \tanh \kappa (\tau -\tau_0) \right]_{,_\theta} , \nonumber
\\ 
\nonumber \\
P(\tau + \Delta \tau) &=& \tau - \ln \left[\, \left| {{Q(\tau),_\theta}
\over {\kappa}}
\right|
\cosh \kappa (\tau -\tau_0) \right], \nonumber \\
\nonumber \\
\pi_P(\tau + \Delta \tau) &=& 1 - |\kappa| \tanh \kappa (\tau - \tau_0)
\end{eqnarray}
from the variation of $H_1$ (where $\xi$, $\kappa$, and $\tau_0$ are
functions of $\theta$) and
\begin{eqnarray}
\label{ber-newh2soln}
P(\tau + \Delta \tau ) &=& P(\tau), \nonumber \\
\nonumber \\
\pi_Q(\tau + \Delta \tau) &=& \pi_Q(\tau), \nonumber \\
\nonumber \\
\pi_P(\tau + \Delta \tau) &=& \pi_P(\tau)  + \pi_Q^2(\tau) \, e^{-2
P(\tau)}\Delta \tau + {\textstyle{1
\over 2}} e^{-2 \tau} \left( 1 - e^{-2 \Delta \tau} \right) P(\tau),_{\theta
\theta} \ , \nonumber \\
\nonumber \\
Q(\tau + \Delta \tau) &=& Q(\tau) + \pi_Q(\tau) \,e^{-2P(\tau)} \Delta \tau
\end{eqnarray}
from the variation of $H_2$. Application of this algorithm is in progress.

From the Gowdy test case, we learn that:

(1) Since the singularity is AVTD, $H_K$ dominates $H_V$ asymptotically so our
(current) algorithm is very accurate.

(2) Non-linear terms in the wave equations act as potentials. In the Gowdy
case, they drive the system to the AVTD regime as $\tau \to \infty$ with $0 \le
v < 1$, where the potentials permanently die out.

(3) Non-generic points where $\pi_Q = 0$ or $Q,_\theta = 0$ lead to the growth
of spiky features in $P$ and $Q$.

(4) Spiky features appear narrower with finer spatial resolution.

\section{$U(1)$ Symmetric Cosmologies}
Given our understanding of the Gowdy model, we can move to spatially
inhomogeneous cosmologies with one Killing field rather than two, retaining a
$U(1)$ symmetry on $T^3 \times R$ \cite{ber-Moncrief86}. These models can be
described by five degrees of freedom \{$\varphi$, $\omega$, $\Lambda$, $z$, $x$
\} and their respective conjugate momenta \{$p$, $r$, $p_\Lambda$, $p_z$,
$p_x$\} which are functions of spatial variables $u$ and $v$ and time $\tau$.
Einstein's equations may be obtained by variation of \cite{ber-bkbvm,ber-bkbvm2}
\begin{eqnarray}
\label{ber-Hu1}
H &=& \oint \oint du \kern 1pt dv \,{\cal H} \nonumber \\
\nonumber \\
&=& \oint \oint du \kern 1pt dv \left( {\textstyle{1 \over 8}}p_z^2\,+\,
{\textstyle {1
\over 2}} e^{4z}p_x^2\,+\,{\textstyle{1 \over 8}}p^2\,+\,{\textstyle{1 \over
2}}e^{4\varphi }r^2\,-\,{\textstyle{1
\over 2}}p_\Lambda ^2+2p_\Lambda  \right) \nonumber \\
\nonumber \\
&& \,+\,e^{-2\tau } \oint \oint du \kern 1pt 
dv \left\{  \left( {e^\Lambda e^{ab}} \right) ,_{ab}\,-\, \left( {e^\Lambda
e^{ab}}
\right) ,_a\Lambda ,_b\,+\,e^\Lambda  \right. \left[  \left( {e^{-2z}}
\right) ,_u x,_v\,-\, \left( {e^{-2z}} \right) ,_v x,_u \right] \nonumber \\
\nonumber \\
&& \left. +\,2e^\Lambda e^{ab}\varphi ,_a\varphi ,_b\,+\,{\textstyle{1 \over 2}}
e^\Lambda e^{-4\varphi }e^{ab}\omega ,_a\omega ,_b \right\} \nonumber \\
\nonumber \\
&=& H_K+H_V = \oint \oint du \kern 1pt dv \, {\cal H}_K +\oint \oint du \kern
1pt dv \,V.
\end{eqnarray}
Here $\varphi$ and $\omega$ are analogous to $P$ and $Q$ while $\Lambda$, $x$,
$z$ describe the metric $\tilde e_{ab}$ in the $u$-$v$ plane perpendicular to
the symmetry direction with
\begin{equation}
\label{ber-confmetr}
\tilde e_{ab}=e^{2\Lambda }e_{ab}={\textstyle{1 \over 2}}e^{2\Lambda }\left(
{\matrix{{e^{2z}+e^{-2z}(1+x)^2}\ &\ {e^{2z}+e^{-2z}(x^2-1)}\cr
\cr
{e^{2z}+e^{-2z}(x^2-1)}\ &\ {e^{2z}+e^{-2z}(1-x)^2}\cr }} \right).
\end{equation}
This model is sufficiently generic that local Mixmaster dynamics is allowed. The
$U(1)$ Hamiltonian (\ref{ber-Hu1}) has the standard symplectic form. Note that
$H_K$ consists of two Gowdy-like $H_K$'s and a free particle term so that $H_K$
is exactly solvable. $H_V$ is also (again trivially) exactly solvable although
spatial differencing must be performed with care. In two spatial dimensions,
there are a variety of ways to represent derivatives to a given order of
accuracy. We currently use a scheme provided by Norton
\cite{ber-norton} to minimize the growth of short wavelength modes.

Unlike the Gowdy model, the constraints and initial value problems must be
considered. We find 
\begin{equation}
\label{ber-H0}
{\cal H}^0 = {\cal H} - 2 p_\Lambda = 0
\end{equation}
and
\begin{eqnarray}
\label{ber-Hu}
{\cal H}^u&=&p_z\,z,_u\,+\,p_x\,x,_u\,+\,p_\Lambda \,\Lambda ,_u\,-\,p_\Lambda
,_u\,+\,p\,\varphi ,_u\,+\,r\,\omega ,_u \nonumber \\
\nonumber \\
& &+{\textstyle{1 \over 2}}\left\{ {\left[ {e^{4z}-(1+x)^2}
\right]p_x-(1+x)p_z} \right\},_v \nonumber \\
\nonumber \\
  & &-{\textstyle{1 \over 2}}\left\{ {\left[ {e^{4z}+(1-x^2)} \right]p_x-x\kern
1pt p_z} \right\},_u=0,
\end{eqnarray}

\begin{eqnarray}
\label{ber-hv}
{\cal H}^v&=&p_z\,z,_v\,+\,p_x\,x,_v\,+\,p_\Lambda \,\Lambda ,_v\,-\,p_\Lambda
,_v\,+\,p\,\varphi ,_v\,+\,r\,\omega ,_v \nonumber \\
\nonumber \\
& &-{\textstyle{1 \over 2}}\left\{ {\left[ {e^{4z}-(1-x)^2}
\right]p_x+(1-x)p_z} \right\},_u \nonumber \\
\nonumber \\
 & &+{\textstyle{1 \over 2}}\left\{ {\left[ {e^{4z}+(1-x^2)} \right]p_x-x\kern
1pt p_z} \right\},_v=0.
\end{eqnarray}
While a general solution to the initial value problem is not known, we use the
particular solution obtained as follows: To solve the momentum constraints
(\ref{ber-Hu}) and (\ref{ber-hv}) set $p_x = p_z = \varphi,_a = \omega,_a = 0$
to leave
$p_\Lambda \, \Lambda,_a -\, p_\Lambda,_a = 0$ which may be satisfied by
requiring $p_\Lambda = c \,e^\Lambda$. For sufficiently large $c$, the
Hamiltonian constraint may be solved algebraically for either $p$ or $r$. In
general, this leaves as free data the four functions $x$, $z$, $\Lambda$, and
either $r$ or $p$. Since there are four free functions at each spatial point,
we expect generic behavior. 
\begin{figure}[bth]
\begin{center}
\setlength{\unitlength}{1cm}
\makebox[11.7cm]{\psfig{file=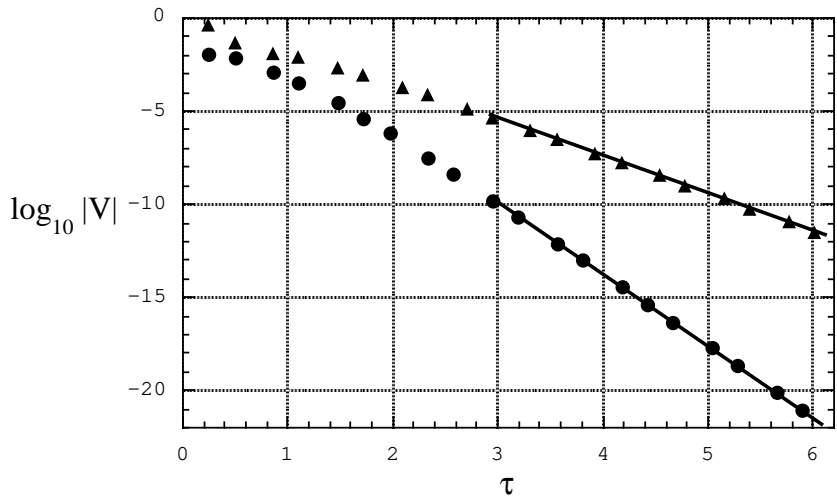,width=11cm}}
\caption[V evolution polarized]
{\protect \label{ber-fig6}
Evolution of $\log_{10}|V|$ vs $\tau$ for a polarized $U(1)$ cosmology at two
representative spatial points. The solid lines are exponential fits.}
\end{center}
\end{figure}
\begin{figure}[bth]
\begin{center}
\setlength{\unitlength}{1cm}
\makebox[11.7cm]{\psfig{file=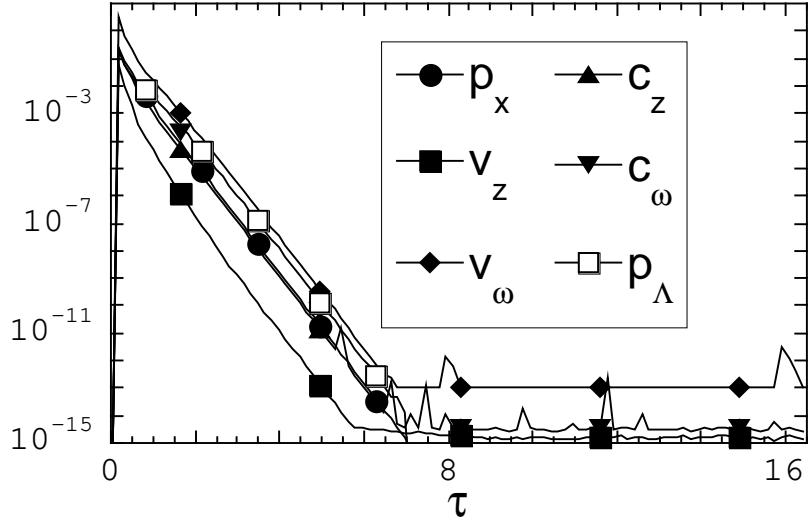,width=11cm}}
\caption[AVTD constants]
{\protect \label{ber-fig7}
The {\bf maximum} value of the {\bf change} with time over the spatial grid of
AVTD regime constants for a polarized $U(1)$ symmetric model. Here $v_z =
\sqrt{p_z^2/8 +e^{4 z} p_x^2/2}$, $v_\omega =\sqrt{ p^2/8 + e^{4
\varphi}r^2/2}$,
$c_z = p_z/2 + p_x \,x$, and $c_\omega = p/2+r\, \omega$. Exponential decay is
observed until the maximum values of the changes reach the level of machine
precision.}
\end{center}
\end{figure}

As a first case, we consider polarized $U(1)$ models obtained by setting
\begin{equation}
\label{ber-u1pol}
r = \omega = 0.
\end{equation}
This condition is preserved numerically as well as analytically. It has been
conjectured \cite{ber-bm2} that polarized $U(1)$ models are AVTD. This is
reasonable because the Mixmaster potential-like term 
\begin{equation}
V_{\nabla \omega} =
e^{-2\tau} e^\Lambda e^{ab} e^{-4 \varphi} \omega,_a \omega,_b
\end{equation}
is absent. Other
spatial derivative terms decay as $e^{\Lambda -2 \tau - 2z}$ for the expected
AVTD limits of the variables as $\tau \to \infty$ \cite{ber-bkbvm2}: 
\begin{eqnarray}
z &\to& -v_z\tau \quad, \quad x \to
x_0 \quad , \quad p_z \to -v_z \quad ,\nonumber \\ 
\nonumber \\
p_x &\to& p_x^0 \quad, \quad \varphi
\to -v_\varphi\tau \quad, \quad p \to -v_\varphi \quad , \nonumber \\
\nonumber \\ 
\Lambda &\to&
\Lambda_0 \,+\,(2 - p_\Lambda^0)\tau \quad , \quad p_\Lambda \to p_\Lambda^0,
\end{eqnarray}
so that an AVTD singularity is consistent.
\begin{figure}[bth]
\begin{center}
\setlength{\unitlength}{1cm}
\makebox[11.7cm]{\psfig{file=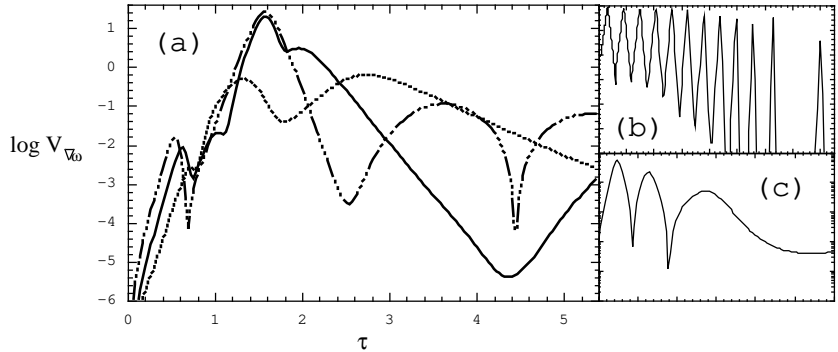,width=11cm}}
\caption[generic V]
{\protect \label{ber-fig8}
(a) $\log_{10}V$ vs $\tau$ for three representative spatial points of a generic
$U(1)$ symmetric cosmology. (b) The analogous quantity for a typical Mixmaster
evolution. (c) The analogous quantity for a Gowdy simulation.}
\end{center}
\end{figure}
\begin{figure}[bth]
\begin{center}
\setlength{\unitlength}{1cm}
\makebox[11.7cm]{\psfig{file=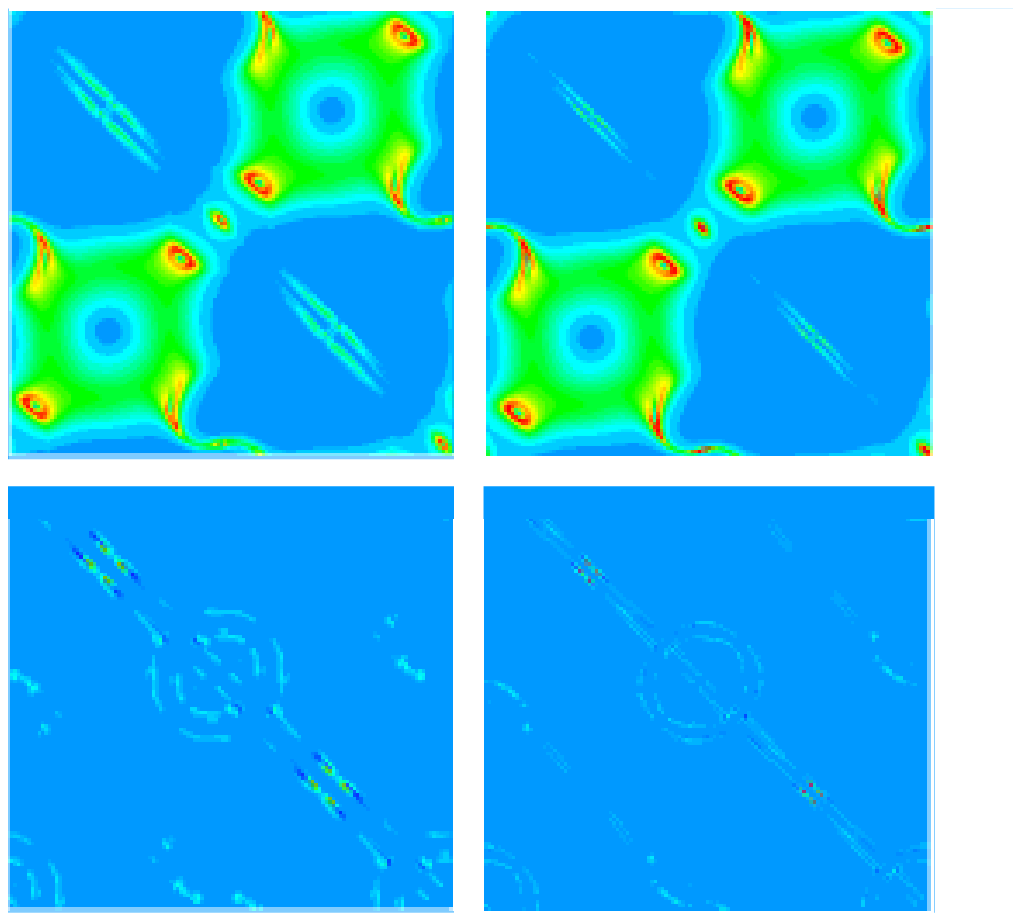,width=9cm}}
\caption[resolution]
{\protect \label{ber-fig9}
Movie frames of $V(u,v)$ for a generic $U(1)$ symmetric cosmology for two fixed
values of $\tau$. The upper frames precede the lower ones. The spatial
coordinates $u$ and $v$ run from $0$ to $2
\pi$ in both directions. The frames on the left are from a simulation with
$128^2$ spatial grid points while the ones on the right have $256^2$ spatial
grid points. Both simulations have the same initial data. Note that there are
features independent of spatial resolution as well as those that depend on the
resolution. The background shade of gray indicates regions where $V \approx 0$.}
\end{center}
\end{figure}

Fig.~\ref{ber-fig6} shows $\log_{10}|V|$ vs $\tau$ for typical spatial points.
Thus we see the expected exponetial decay. In Fig.~\ref{ber-fig7}, we see that
the maximum values of the changes with time of the quantitites expected to be
constant in an AVTD regime also decay exponentially as expected. These results
will be discussed in detail elsewhere \cite{ber-bkbvm2}.

We emphasize here that the polarized $U(1)$ models present {\it no} numerical
difficulties. The absence of $V_{\nabla \omega}$ means that spiky features do
not develop. The situation unfortunately changes for generic (unpolarized)
$U(1)$ models. It appears that consistency arguments which suggest an AVTD
singularity in polarized $U(1)$ models and restrict $v$ to $[0,1]$ in the Gowdy
models fail in generic $U(1)$ models. This suggests that $V_{\nabla \omega}$
will always grow exponentially if the system tries to be AVTD producing a
Mixmaster-like bounce. A bounce in the opposite direction will come, as in
Gowdy models, from terms in $H_K$. This bouncing could continue indefinitely.

Unfortunately, the bounces off $V_{\nabla \omega}$ probably cause numerical
instabilities that limit the duration of the simulations. Spatial averaging has
been used to improve stability but it is known to produce numerical artifacts.
Typical evolutions of $V_{\nabla \omega}$ at single spatial grid points are
shown in Fig.~\ref{ber-fig8} and compared to similar quantities in Mixmaster and
Gowdy models and, of course, to Fig.~\ref{ber-fig6}. While generic $U(1)$
models are clearly different from polarized ones, it is not clear yet whether
the observed bounces are Mixmaster-like or will eventually die out as in Gowdy
models. Recall that Mixmaster universes are very close to the AVTD Kasner
solution between bounces. One is also concerned about numerical artifacts
although it is not clear that standard tests are helpful. Fig.~\ref{ber-fig9}
shows two frames from a movie of $V(u,v)$ vs
$\tau$ for two different spatial resolutions. Features are narrower on the
finer grid. While this usually indicates artifacts, we recall that this is
precisely what happens in Gowdy models where the resolution dependence is well
understood.

\section{Future Directions}
In the search for the nature of the generic cosmological singularity, we have
obtained convincing evidence that both the Gowdy universes and the polarized
$U(1)$ symmetric cosmologies have AVTD singularities. These results are found
in simulations which present no numerical difficulties. In contrast, we cannot
draw definite conclusions for generic $U(1)$ models except to say that we have
not found evidence that the singularities are AVTD everywhere.

Presumably, numerical difficulties in generic $U(1)$ models are due to
Gowdy-like spiky features (which are less easy to represent accurately in two
spatial dimensions). It is possible that the new Mixmaster algorithm
\cite{ber-bkb96} which can be adapted to the Gowdy model will help in the
generic
$U(1)$ case where it can also be implemented. The hope is that better treatment
of the bounces would give better local treatment of spiky features that arise
from them. 

While we solve the constraints initially in $U(1)$ models, we do nothing to
preserve them thereafter. In the polarized case, they remain acceptably small
(and in fact converge to zero with increasing spatial resolution). We have
learned from the Mixmaster case that one must preserve the constraints
\cite{ber-bkb96}. In fact, it is the kinetic part of the Hamiltonian constraint
which restricts the exponential factor in $V_{\nabla \omega}$. An error in the
constraints could give the argument of the exponential the wrong sign leading
to the observation of qualitatively wrong behavior. By starting closer to the
singularity, we can supress some of the numerical instabilities and study this
controlling exponential. Studies of this type have provided some evidence that
it is essential to solve the constraints. Work on implementing a constraint
solver is in progress.

\section*{Acknowledgements}
B.K.B. and V.M. would like to thank the Albert Einstein Institute at Potsdam
for hospitality. B.K.B. would also like to thank the Institute of Geophysics
and Planetary Physics of Lawrence Livermore National Laboratory for
hospitality. This work was supported in part by National Science Foundation
Grants PHY9507313 and PHY9722039 to Oakland University and PHY9503133 to Yale
University. Computations were performed at the National Center for
Supercomputing Applications (University of Illinois).

\end{document}